# Irreducible Green Functions Method applied to nanoscopic systems

Grzegorz Górski[*]

*Faculty of Mathematics and Natural Sciences, University of Rzeszów*
*ul. Pigonia 1, 35-959 Rzeszów, Poland*
*\*Corresponding author's e-mail: ggorski@ur.edu.pl*

The equation of motion method (EOM) for Green functions is one of the tools used in the analysis of quantum dot system coupled with the metallic and superconducting leads. We investigate modified EOM, based on differentiation of double-time temperature dependent Green functions both over the primary time $t$ and secondary time $t'$. Our EOM approach allows us to obtain the Abrikosov-Suhl resonance in the particle-hole symmetric case and also in the asymmetric cases. We will apply the irreducible Green functions technique to analyses of the EOM applied to the dot system. This method give a workable decoupling scheme breaking the infinite set of Green function equations. We apply this technique to calculate the density of the states and the differential conductance of single-level quantum dot with Coulomb repulsion attached to one metallic and one superconducting leads (N-QD-SC). Our results are compared with the previous calculations.

## 1. Introduction

Electronic transport through the quantum dots is a very interesting effect which at present is investigated experimentally and theoretically. One of the important examples is the system in which the quantum dot is coupled with one metallic and one superconducting leads (N-QD-SC). In this system there is an interplay between the Andreev refraction and the Kondo effect. Transition between those two effects depends strongly on the Coulomb interaction. For the weak interaction we observe superconducting singlet state with two Andreev resonances in the spectral density. At large Coulomb interaction the Kondo singlet is preferred in the spectral density and the Andreev resonances overlap and become the central Kondo peak. The competition between these two singlet states can be investigated by the linear conductance measurements. In the crossover region the zero bias conductance [1] has the maximum. In the Kondo effect region the conductance decreases.

Different methods are used for theoretical analysis of the Coulomb correlation in N-QD-SC system: numerical renormalization group (NRG) [1], modified perturbation theory (MPT) [2-4], equation of motion approach [5, 6]. The weak point of standard equation of motion (EOM) approach is that it does not describe the Kondo state in the particle-hole symmetric case. We use modified (EOM) approach [7] in which we differentiate Green functions over both time variables. This differs from the standard EOM method where the time derivative was taken only over primary time variable. Such an approach allows to describe correctly the Kondo state in particle-hole symmetric case but also in the asymmetric case. Using this method we calculated the spectral density of states and the linear conductance of the N-QD-SC system. The results are compared with the results of NRG method [1, 8].

## 2. The model

We analyze the system that is build out of quantum dot connected to one metallic lead and one superconducting lead. Hamiltonian of this model has the form

$$H = \sum_\sigma \varepsilon_d n_{d\sigma} + U n_{d\uparrow} n_{d\downarrow} + \sum_{\substack{\mathbf{k}\sigma \\ \alpha=N,S}} (\varepsilon_{\mathbf{k}\alpha} - \mu_\alpha) n_{\mathbf{k}\alpha\sigma} + \sum_{\substack{\mathbf{k}\sigma \\ \alpha=N,S}} \left(V_{\mathbf{k}\alpha} d_\sigma^+ c_{\mathbf{k}\alpha\sigma} + h.c.\right)$$
$$-\Delta \sum_{\mathbf{k}} \left(c_{\mathbf{k}S\uparrow}^+ c_{-\mathbf{k}S\downarrow}^+ + c_{-\mathbf{k}S\downarrow} c_{\mathbf{k}S\uparrow}\right) \qquad (1)$$

where $d_\sigma^+(d_\sigma)$ are the creation (annihilation) operators for the dot electron with spin $\sigma$, $c_{\mathbf{k}\alpha\sigma}^+(c_{\mathbf{k}\alpha\sigma})$ where $\alpha = N,S$ are the creation (annihilation) operators for the electron in the normal (N) or superconducting (S) lead, $\varepsilon_{\mathbf{k}\alpha}$ is the energy dispersion of $\alpha$ lead, $\mu_\alpha$ is the chemical potential of $\alpha$ lead, $\varepsilon_d$ is the dot energy, $U$ is the on-site Coulomb interaction between electrons on the dot, and $V_{\mathbf{k}\alpha}$ is the coupling between the $\alpha$ lead and the dot. We assume that the superconducting lead is well described by the BCS theory with a superconducting gap $\Delta$.

We are interested in the physics of the Andreev reflection therefore for simplicity we will assume that $\Delta \to \infty$. In effect eq. (1) is reduced to the following effective Hamiltonian [9]

$$H = \sum_\sigma \varepsilon_d n_{d\sigma} + U n_{d\uparrow} n_{d\downarrow} + \sum_{\mathbf{k}\sigma} (\varepsilon_{\mathbf{k}N} - \mu_N) n_{\mathbf{k}N\sigma} + \sum_{\mathbf{k}\sigma} \left(V_{\mathbf{k}N} d_\sigma^+ c_{\mathbf{k}N\sigma} + h.c.\right) - \Delta_d \left(d_\uparrow^+ d_\downarrow^+ + d_{-\downarrow} d_\uparrow\right) \quad (2)$$

where $\Delta_d = \Gamma_S$ denotes the effective onsite superconducting gap at the QD. The parameter $\Gamma_S = \pi \sum_\mathbf{k} |V_{\mathbf{k}S}|^2 \delta(\varepsilon - \varepsilon_{\mathbf{k}S})$ describe the resonance strength between the QD and the superconducting lead. In the broad band limit we assume that it is a constant value.

We are searching for the Green's function in the Nambu space. In the energy representation we can write it as:

$$\hat{\mathbf{G}}_d(\varepsilon) = \begin{bmatrix} \langle\langle d_\uparrow; d_\uparrow^+ \rangle\rangle_\varepsilon & \langle\langle d_\uparrow; d_\downarrow \rangle\rangle_\varepsilon \\ \langle\langle d_\downarrow^+; d_\uparrow^+ \rangle\rangle_\varepsilon & \langle\langle d_\downarrow^+; d_\downarrow \rangle\rangle_\varepsilon \end{bmatrix} \qquad (3)$$

In our analysis we will use the equation of motion technique for the double-time Green's functions. In general the EOM for Green's functions is obtained by differentiation over primary time ($t$), what after Fourier transform brings the following equation

$$\varepsilon \langle\langle A; B \rangle\rangle_\varepsilon = \langle [A,B]_+ \rangle + \langle\langle [A,H]_-; B \rangle\rangle_\varepsilon . \qquad (4)$$

In this form eq. (4) is widely used in analysis of nanoscopic systems (see e.g. [5, 10]). It allowed for obtaining the Abrikosov-Suhl resonance, but only outside of the particle-hole symmetric system. For $n_d = 1$ in this approach the Abrikosov-Suhl peak disappear. Less frequently used method in solving the EOM is differentiating the double-time Green's function over the second time ($t'$). In the energy representation this leads to the following equation [7]:

$$-\varepsilon \langle\langle A; B \rangle\rangle_\varepsilon = -\langle [A,B]_+ \rangle + \langle\langle A; [B,H]_- \rangle\rangle_\varepsilon . \qquad (5)$$

In our approach we will use both forms of equation of motion. To analyze the higher order Green's functions ($\langle\langle [A,H]_-; B \rangle\rangle_\varepsilon$ and $\langle\langle A; [B,H]_- \rangle\rangle_\varepsilon$) we will apply the irreducible Green's functions technique [11]. By the help of eqs. (4) and (5) we obtain:

$$\hat{\mathbf{g}}_d^{HF-1} \hat{\mathbf{G}}_d = \hat{\mathbf{I}} + U^2 \hat{\boldsymbol{\tau}}_3^{ir} \hat{\boldsymbol{\Gamma}}_d^{(2)ir} \hat{\boldsymbol{\tau}}_3 \hat{\mathbf{g}}_d^{HF} \qquad (6)$$

where $\hat{\tau}_i$ is the Pauli matrix, $\hat{g}_d^{HF}$ is the QD Hartree-Fock Green's function in the Nambu space

$$\hat{g}_d^{HF-1} = \begin{bmatrix} \varepsilon - \varepsilon_d + i\Gamma_N - U\langle n_{d\downarrow}\rangle & \Delta_d - U\langle d_\downarrow d_\uparrow\rangle \\ \Delta_d - U\langle d_\uparrow^+ d_\downarrow^+\rangle & \varepsilon + \varepsilon_d - i\Gamma_N + U\langle n_{d\uparrow}\rangle \end{bmatrix}, \quad (7)$$

the expression $^{ir}\hat{\Gamma}_d^{(2)ir}$ is irreducible Green's functions

$$^{ir}\hat{\Gamma}_d^{(2)ir} = \begin{bmatrix} ^{ir}\langle\langle \hat{n}_{d\downarrow}d_\uparrow; n_{d\downarrow} d_\uparrow^+\rangle\rangle_\varepsilon^{ir} & ^{ir}\langle\langle \hat{n}_{d\downarrow}d_\uparrow; \hat{n}_{d\uparrow}d_\downarrow\rangle\rangle_\varepsilon^{ir} \\ ^{ir}\langle\langle \hat{n}_{d\uparrow}d_\downarrow^+; n_{d\downarrow} d_\uparrow^+\rangle\rangle_\varepsilon^{ir} & ^{ir}\langle\langle \hat{n}_{d\uparrow}d_\downarrow^+; \hat{n}_{d\uparrow}d_\downarrow\rangle\rangle_\varepsilon^{ir} \end{bmatrix}, \quad (8)$$

and $\Gamma_N = \pi\sum_{\mathbf{k}}|V_{\mathbf{k}N}|^2 \delta(\varepsilon - \varepsilon_{\mathbf{k}N})$ represents the resonance strength between the QD and the metallic lead. Using the Dyson's relations and dividing the self-energy: $\hat{\Sigma}_U = \hat{\Sigma}_U^{HF} + \hat{\Sigma}'_U$, into the first order term $\hat{\Sigma}_U^{HF}$ and the remaining higher order part ($\hat{\Sigma}'_U$) we obtain:

$$\hat{\Sigma}_U^{HF} = \begin{bmatrix} U\langle n_{d\downarrow}\rangle & U\langle d_\downarrow d_\uparrow\rangle \\ U\langle d_\uparrow^+ d_\downarrow^+\rangle & -U\langle n_{d\uparrow}\rangle \end{bmatrix} \quad (9)$$

and

$$\hat{\Sigma}'_U = \left[\hat{\mathbf{I}} + U^2\hat{\tau}_3\,^{ir}\hat{\Gamma}_d^{(2)ir}\hat{\tau}_3\hat{g}_d^{HF}\right]^{-1} U^2\hat{\tau}_3\,^{ir}\hat{\Gamma}_d^{(2)ir}\hat{\tau}_3 . \quad (10)$$

The irreducible Green's functions $^{ir}\hat{\Gamma}_d^{(2)ir}$ defined by eq. (8) cannot be reduced to the lower order Green's function by any kind of decoupling. Using spectral theorem we obtain the approximate expression for $^{ir}\hat{\Gamma}_d^{(2)ir}$:

$$^{ir}\hat{\Gamma}_d^{(2)ir}(\varepsilon) = -\frac{1}{\pi}\int_{-\infty}^{\infty}\frac{d\varepsilon'}{\varepsilon - \varepsilon' + i0^+}\,\text{Im}\,\hat{X}(\varepsilon'). \quad (11)$$

where

$$\text{Im}\,\hat{X}(\varepsilon) = -\int_{-\infty}^{\infty}\left[\Pi_1(\varepsilon+\varepsilon')\hat{\tau}_2\left[\hat{\rho}_{HF}^+(\varepsilon')\right]^T \hat{\tau}_2 + \Pi_2(\varepsilon+\varepsilon')\hat{\tau}_2\left[\hat{\rho}_{HF}^-(\varepsilon')\right]^T \hat{\tau}_2\right]d\varepsilon' \quad (12)$$

$$\Pi_{1(2)}(\varepsilon) = \pi\int_{-\infty}^{\infty}\left[\rho_{HF11}^{-(+)}(\varepsilon')\rho_{HF22}^{-(+)}(\varepsilon-\varepsilon') - \rho_{HF12}^{-(+)}(\varepsilon')\rho_{HF21}^{-(+)}(\varepsilon-\varepsilon')\right]d\varepsilon'. \quad (13)$$

In equations (12) and (13) we have introduced the state densities: $\rho_{ij}^\pm(\varepsilon) = -1/\pi\,\text{Im}\,g_{d,ij}^0(\varepsilon)f^\pm(\varepsilon)$, where $f^\pm(\varepsilon) = \left[1+\exp(\pm\varepsilon/T)\right]^{-1}$ is the particle/hole distribution functions. To calculate functions $g_{d,ij}^0(\varepsilon)$ we used effective dot energy level and effective local pairing potential determined in a self-consistent manner [2].

Similar form of the self-energy for N-QD-SC system was used by Martin-Rodero et. al. [3, 4], and Yamada et.al [2], but in their papers the denominator of the self-energy was introduced by interpolation to the results of the weak and strong coupling limits. In our approach denominator in eq. (10) is derived directly from the corrected two times EOM method.

### 3. Numerical results and discussion

In the numerical analysis we will concentrate on the equilibrium system where $\mu_S = \mu_N = 0$. The unit of energy is the coupling parameter between dot and metallic lead $\Gamma_N = 1$. In Fig. 1 (a) we present the energy dependence of the correlated quantum dot spectral density, $\rho_d(\varepsilon) = -\operatorname{Im} G_{11}(\varepsilon)/\pi$, for different values of Coulomb interaction in the particle-hole symmetric coupling case. At small values of the Coulomb interaction we obtain two pronounced Andreev quasiparticle peaks. This situation points to the existence of the superconducting singlet state. Increase of the Coulomb interaction favors Kondo effect and as the result we observe crossover between Copper-pairing singlet state and the Kondo singlet state. In the spectral density it causes renormalization of the Andreev levels and their approach to the Fermi level. Finally both states become one resonance state typical for the the Kondo effect. In addition to this dominant resonance there are four other resonances representing the electron and hole components of the excitations to the upper and lower atomic levels [6, 8]. These results are comparable with the numerical results obtained by the renormalization group (NRG) method [8].

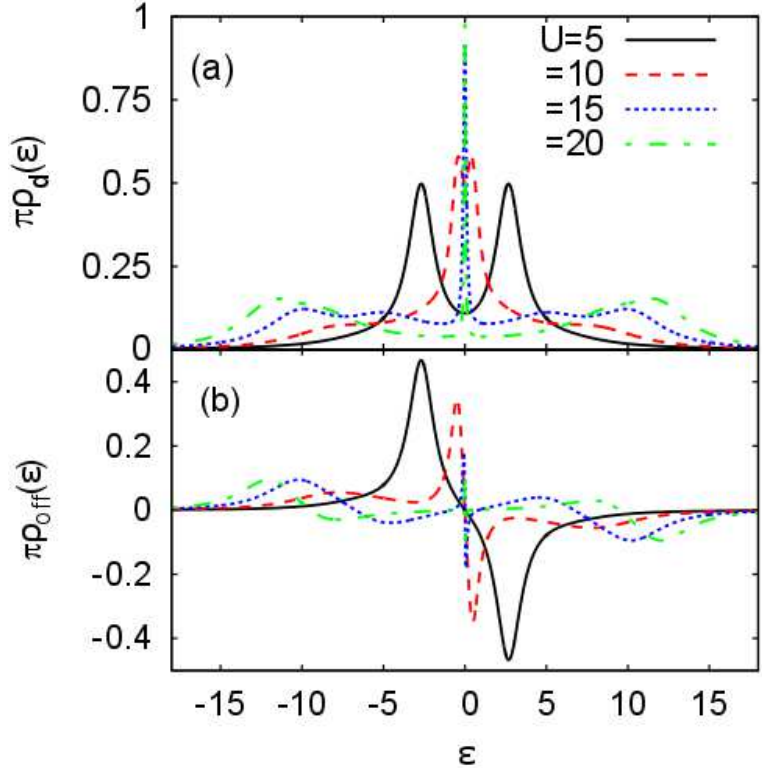

Fig. 1 (a) Spectral density $\rho_d(\varepsilon)$ of correlated quantum dot plotted as a function of energy for different values of Coulomb interaction, $\Gamma_N = 1$, $\Gamma_S = 5$ and particle hole symmetric case. (b) Anomalous spectral density $\rho_{off}(\varepsilon)$ of correlated quantum dot.

In Fig. 1(b) we show the anomalous spectral density $\rho_{off}(\varepsilon) = -\operatorname{Im} G_{21}(\varepsilon)/\pi$ for different values of $U$. With the increase of $U$ value of $\rho_{off}(\varepsilon)$ decreases. Further increase of $U$ leads to the sign change of $\rho_{off}(\varepsilon)$. At $U<10$ and $\varepsilon>0$ we have negative $\rho_{off}(\varepsilon)$. At $U>10$ we obtain positive $\rho_{off}(\varepsilon)$ at some energies. Six extreme points in the dependence of $\rho_{off}(\varepsilon)$ (seen particularly well at large $U>10$) correspond to the resonance levels in the spectral density, where two states close to the Fermi energy merge into one Kondo peak [8]. Sign change in the anomalous spectral density is the criterion of the transition from the superconducting singlet- Kondo state to the single (doublet) state [12]. In crossover region the anomalous spectral density disappears but the spectral density $\rho_d(\varepsilon)$ rapidly grows (see Fig. 1(a)). Similar behavior of $\rho_d(\varepsilon)$ and $\rho_{off}(\varepsilon)$ near the transition point was observed by Zitko et. al [12] in the systems with finite superconducting energy gap.

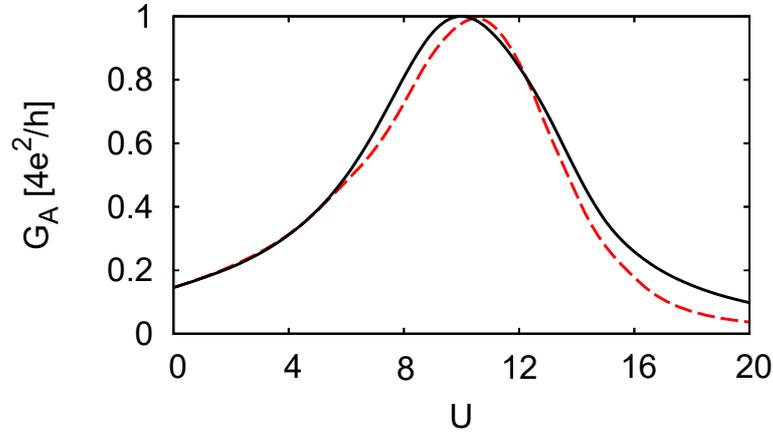

Fig. 2 Andreev linear conductance as a function of the Coulomb interaction $U$ for $\Gamma_N = 1$ and $\Gamma_S = 5$ calculated in our approach (solid line) and from NRG method (dashed line) [1].

In Fig. 2 we show Andreev linear conductance as a function of the Coulomb interaction for $\Gamma_N = 1$ and $\Gamma_S = 5$. In calculations we have used expression derived by Baranski and Domanski [6]. At small $U<10$ we observe increase of Andreev linear conductance with the increase of $U$. In the crossover region ($U=10$) the linear conductance reaches its maximum value of $4e^2/h$. Further increase of $U$ causes transition to the Kondo physics and as the result $G_A$ decreases. For comparison we present also the results from numerical renormalization group method [1]. One can see that there is good agreement between those results.

Reassuming the method of modified EOM applied to N-QD-SC system allows to analyze mutual interplay between the proximity induced pairing and the Kondo-type correlations. Results obtained for DOS and for the Andreev linear conductance are in agreement with the other numerical methods. Our method based on the analytical approach describes well transition from superconducting to singlet- Kondo state in the broad range of the Coulomb correlations

**Acknowledgments**

This work was done with the support from Faculty of Mathematics and Natural Sciences University of Rzeszów within the project no. WMP/GD-06/2015.

**References**
[1] Y. Tanaka, A. Kawakami, A. Oguri, *J. Phys. Soc. Jpn.* **76**, 074701 (2007). doi: 10.1143/JPSJ.074701
[2] Y. Yamada, Y. Tanaka, N. Kawakami, *Phys. Rev. B* **84**, 075484 (2011). doi: 10.1103/PhysRevB.84.075484
[3] A. Martín-Rodero, A. Levy Yeyati, *Adv. Phys.* **60**, 899 (2011). doi: 10.1080/00018732.2011.624266
[4] J. C. Cuevas, A. Levy Yeyati, A. Martín-Rodero, *Phys. Rev. B* **63**, 094515 (2001). doi: 10.1103/PhysRevB.63.094515
[5] J. Barański, T. Domański, *Phys. Rev. B* **84**, 195424 (2011). doi: 10.1103/PhysRevB.84.195424
[6] J. Barański, T. Domański, *J. Phys. Condens. Matter.* **25**, 435305 (2013). doi:10.1088/0953-8984/25/43/435305
[7] G. Górski, J. Mizia, K. Kucab, *Physica E* **73**, 76 (2015). doi:10.1016/j.physe.2015.05.021
[8] A. Oguri, Y. Tanaka, J. Bauer, *Phys. Rev. B* **87**, 075432 (2013). doi: 10.1103/PhysRevB.87.075432
[9] A.V. Rozhkov and D.P. Arovas, *Phys. Rev. B* **62**, 6687 (2000). doi: 10.1103/PhysRevB.62.6687
[10] P. Trocha, J. Barnaś, *Phys. Rev. B* **76** (2007) 165432 doi: 10.1103/PhysRevB.76.165432
[11] K.I. Wysokiński, A.L. Kuzemsky, *J Low Temp. Phys.* **52,** 81 (1983). doi: 10.1007/BF00681267
[12] R. Žitko, J. S. Lim, R. López, R. Aguado, *Phys. Rev. B* **91**, 045441 (2015). doi: 10.1103/PhysRevB.91.045441